\documentclass[aps,prb,reprint,groupedaddress,amsfonts,amssymb,amsmath]{revtex4-1}

\usepackage{lineno,hyperref,amsmath,graphicx}

\begin{document}

\title{Local distributions of the 1D dilute Ising model}

\author{Yu.D. Panov}
\email[]{yuri.panov@urfu.ru}
\affiliation{Ural Federal University, Ekaterinburg, Russia}

\begin{abstract}
The local distributions of the one-dimensional dilute annealed Ising model with charged impurities are studied. 
Explicit expressions are obtained for the pair distribution functions and correlation lengths, and their low-temperature asymptotic behavior is explored depending on the concentration of impurities. 
For a more detailed consideration of the ordering processes, we study local distributions. 
Based on the Markov property of the dilute Ising chain, 
we obtain an explicit expression for the probability of any finite sequence 
and find a geometric probability distribution for the lengths of sequences consisting of repeating blocks. 
An analysis of distributions shows that the critical behavior of the spin correlation length is defined by ferromagnetic or antiferromagnetic sequences, 
while the critical behavior of the impurity correlation length is defined by the sequences of impurities or by the charge-ordered sequences. 
For the dilute Ising chain, there are no other repeating sequences whose mean length diverges at zero temperature. 
While both the spin correlation and the impurity correlation lengths can diverge only at zero temperature, 
the ordering processes result in a maximum of the specific heat at finite temperature defined by the maximum rate of change of the impurity-spin pairs concentration. 
A simple approximate equation is found for this temperature. 
We show that the non-ordered dilute Ising chains correspond to the regular Markov chains, while various orderings generate the irregular Markov chains of different types. 
\end{abstract}

\keywords{dilute Ising chain \sep 
correlation functions \sep 
local distributions \sep 
Markov chain}

\maketitle

%%%%%%%%%%%%%%%%%%%%%%%%%%%%%%%%%%%%%%%%%%%%%%%%%%%%%%%%%%%%%%%%%%%%%%%%%%%%%%%%%%%%%%%%%%%%%%%%%%%
%%%%%%%%%%%%%%%%%%%%%%%%%%%%%%%%%%%%%%%%%%%%%%%%%%%%%%%%%%%%%%%%%%%%%%%%%%%%%%%%%%%%%%%%%%%%%%%%%%%
\section{Introduction}

One-dimensional (1D) spin models, including Ising type ones, are convenient objects for testing both the basic concepts of statistical physics and the applicability of new methods. 
To date, exact solutions have been found for various complex models based on the 1D Ising model. 
They includes Blume-Emery-Griffiths model \cite{Wu1978,Thomaz2016,CorreaSilva2016},  
the model with localized Ising-like spins and exchanging electrons \cite{Pereira2008}, 
the models with single-ion anisotropy for the spin-1 chain \cite{Yang2008,Yang2009,DeSouza2014} 
or the mixed spin-1/2 and spin-1 Ising chain \cite{Wu2010,Wu2011,Strecka2011}, 
the Ising-Heisenberg ``decorated'' chains, ladders, and tubes 
\cite{Canova2006,Antonosyan2009,Rojas2011,Galisova2013,Torrico2014,Lisnyi2015,Galisova2015,Rojas2016,Strecka2016,Torrico2018}
and the Ising-Hubbard diamond chain and ladder \cite{Lisnii2011,Sousa2018}.
These systems show many subtle and important phenomena, 
including quantized plateaux in the magnetization curves, quantum entanglement, 
quasi-phases and pseudo-transitions \cite{DeSouza2018},
and describe the properties of real materials, such as the polymeric coordination compounds 
(see Refs. in \cite{Antonosyan2009,Strecka2011,Rojas2016,Strecka2016,Torrico2018,Sousa2018}).
At the same time, the analysis of the conventional 1D Ising model also continues taking into account 
higher spin $S$ values \cite{Suzuki1967}, 
two kinds of spins \cite{Muto1976}, 
the random short- and long-range interactions \cite{Haley1978,Goncalves1998}, 
the next-nearest-neighbour coupling \cite{Fakhri2019} 
and the magnetic field \cite{Kassan-ogly2001,Proshkin2017,Zarubin2019}.

The dilute Ising model is one of the basic ones in the theory of magnetic systems disordered by non-magnetic impurities. 
Undoubtedly, the 1D version of this model has a long history. 
The energy and susceptibility for non-interacting impurities 
was obtained by Katsura and Tsujiyama \cite{Katsura1965} and 
the expressions of thermodynamic functions on the impurities density was given by Kawatra and Kijewski \cite{Kawatra1969}. 
The exact solution and various thermodynamic properties of the dilute Ising chain with interacting impurities 
was found in \cite{Rys1969,Matsubara1973,Termonia1974} 
and in the most general form 
by Balagurov, Vaks and Zaitsev 
in \cite{Balagurov1974}. 
However, despite a long history, the local distributions of spins and non-magnetic interacting impurities in the 1D dilute Ising model have not been described systematically as yet.
The size distribution of clusters is of fundamental interest 
and has been studied intensively in the Ising or similar models 
\cite{Wortis1974,Vaks1975,Harris1975,Binder1976,Marro1983,Thomsen1983,Toral1987,Hu1986,Khn1987,Vavro2001,Campi2003,Yilmaz2005,Simonin2013}. 

In the present paper, we consider the distributions properties of charged non-magnetic impurities and spins for the 1D dilute Ising model. 
From a general point of view, the distributions considered here reveal the reason for the lack of ordering in the 1D dilute Ising model at finite temperature.

The paper is organized as follows. 
In section 2, we obtain the pair distribution functions, the correlation lengths, 
and the probabilities of local distributions in the dilute Ising chain. 
Often local distributions are considered as a way to calculate the thermodynamics of the entire system, so initially, one uses their combinatorial probabilities \cite{Wortis1974,Binder1976,Thomsen1983}. 
Based on the Markov property of the dilute Ising chain, 
we calculate the thermodynamic probabilities, obtaining them from the pair distribution functions. 
Especially, we find the type of the probability distribution for the lengths of sequences of repeated blocks. 
Section 3 describes the features of the ordering processes at low temperatures for various model parameters.
Given the properties of correlation lengths, we examine the distributions for several specific types of local sequences 
and prove the absence of other sequences with diverging mean length. 
In the end, we discuss the properties of the Markov chains, which are generated by the ordered and non-ordered dilute Ising chains.
Conclusions are presented in section 4.

%%%%%%%%%%%%%%%%%%%%%%%%%%%%%%%%%%%%%%%%%%%%%%%%%%%%%%%%%%%%%%%%%%%%%%%%%%%%%%%%%%%%%%%%%%%%%%%%%%%
%%%%%%%%%%%%%%%%%%%%%%%%%%%%%%%%%%%%%%%%%%%%%%%%%%%%%%%%%%%%%%%%%%%%%%%%%%%%%%%%%%%%%%%%%%%%%%%%%%%
\section{Theory}

%%%%%%%%%%%%%%%%%%%%%%%%%%%%%%%%%%%%%%%%%%%%%%%%%%%%%%%%%%%%%%%%%%%%%%%%%%%%%%%%%%%%%%%%%%%%%%%%%%%
\subsection{Pair distribution functions and correlation lengths}

In this work, we consider the dilute Ising chain, which has the Hamiltonian
\begin{equation}
	\mathcal{H} 
	= 
	- J \sum_{j=1}^{N} \sigma_{j} \sigma_{j+1}
	+ V \sum_{j=1}^{N} P_{0,j} P_{0,j+1} 
	- \mu \sum_{j=1}^{N} P_{0,j}
	.
	\label{eq:H}
\end{equation}
Here the pseudospin $\sigma=1$ operator is used, 
where the states of conventional spin doublet and non-magnetic impurity correspond 
to the pseudospin $z$-projections $\sigma = \pm1$ and $\sigma = 0$, respectively,
$J$ is the exchange constant, 
$V$ is the inter-site interaction for impurities,
$P_{0} = 1-\sigma^2$ is the projection operator onto the $\sigma = 0$ state,
and $\mu$ is the chemical potential. 
Further we will assume that nonmagnetic impurities are mobile, which corresponds to the annealed system. 

Detailed information on the state of the thermodynamic system is provided by the pair distribution functions (PDF) 
$\left\langle P_{a,k} \, P_{b,k+l} \right\rangle$, 
where $P_{a,k}$ is the projection operator on one of the basis state, 
$a=\pm,0$  for  $\sigma = \pm1,0$. 
PDF of the 1D dilute Ising model can be calculated by
\begin{equation}
	\left\langle P_{a,k} \, P_{b,k+l} \right\rangle
	= \lim_{N\to\infty} 
	\frac{  \mathop{\mathrm{Tr}} \left(  P_a \mathcal{T}^l P_b \mathcal{T}^{N-l}  \right)  }{  \mathop{\mathrm{Tr}} \left(  \mathcal{T}^{N}  \right)} 
	.
	\label{eq:pdf-def}
\end{equation}
Here $\mathcal{T}$ is the transfer matrix for Hamiltonian (\ref{eq:H}) at $h=0$
\begin{equation}
	\mathcal{T} =
	\left(
	\begin{array}{ccc}
		e^{K}     &  e^{\xi/2}   &  e^{-K}     \\
		e^{\xi/2} &  e^{-W+\xi}  &  e^{\xi/2}  \\
		e^{-K}    &  e^{\xi/2}   &  e^{K} 
	\end{array}
	\right)
	,
\end{equation}
where $K=\beta J$, $W=\beta V$, $\xi=\beta\mu$, $\beta=1/\theta$ and $\theta=k_B T$. 
The activity can be expressed at given concentration of impurities $n$ as
\begin{equation}
	e^{\xi} = 2 \, \frac{ g + m  }{  g - m  } \, e^{W} \cosh K 
	,
	\label{eq:roots}
\end{equation}
where $m = n-1/2$ is the deviation of the concentration of impurities from half-filling and 
\begin{equation}
	g = \left[ m^2 +\left( \frac{1}{4} - m^2 \right) e^{-W} \cosh K \right]^{1/2}
	.
	\label{eq:g}
\end{equation}
From (\ref{eq:pdf-def}) we obtain
\begin{eqnarray}
	\langle P_{0,k} P_{0,k+l} \rangle
	&=&
	\left( \frac{1}{2} + m \right)^2
	+ \left( \frac{1}{4} - m^2 \right) \Lambda_0^{l}  
	,
	\label{eq:P00l}
	\\
	\langle P_{0,k} P_{\pm,k+l} \rangle
	&=&
	\frac{1}{2}
	\left( \frac{1}{4} - m^2 \right)
	\left( 
	1 - \Lambda_0^{l}
	\right)
	,
	\label{eq:P0pl}
	\\
	\langle P_{\pm,k} P_{\pm,k+l} \rangle
	&=&
	\frac{1}{4}
	\bigg[
	\left( \frac{1}{2} - m \right)^2
	+ \left( \frac{1}{4} - m^2 \right) \Lambda_{0}^{l} 
	+{}
	\nonumber \\
	&&
	{}+ \left( \frac{1}{2} - m \right) \Lambda_{\sigma}^{l}
	\bigg]
	,
	\label{eq:Pppl}
	\\
	\langle P_{\pm,k} P_{\mp,k+l} \rangle
	&=&
	\frac{1}{4}
	\bigg[
	\left( \frac{1}{2} - m \right)^2
	+ \left( \frac{1}{4} - m^2 \right) \Lambda_{0}^{l} 
	-{}
	\nonumber \\
	&&
	{}- \left( \frac{1}{2} - m \right)	\Lambda_{\sigma}^{l}
	\bigg]
	,
	\label{eq:Ppml}
\end{eqnarray}
where 
\begin{equation}
	\Lambda_0 = \frac{ 2g - 1 }{ 2g + 1 }
	,\qquad
	\Lambda_{\sigma} = \frac{ g - m }{ g + \frac{1}{2} } \, \tanh K
	.
\end{equation}
In all cases, we have 
\begin{equation}
	\left\langle P_{a,k} \, P_{b,k+l} \right\rangle
	= \left\langle P_{a} \right\rangle  \left\langle  P_{b} \right\rangle
	+ K_{ab}(l)
	,
\end{equation}
where $\left\langle P_{0}\right\rangle = \frac{1}{2} + m$, 
$\left\langle P_{\pm}\right\rangle = \frac{1}{2}\left(\frac{1}{2} - m\right)$, 
and $K_{ab}(l)$ is the correlation function for the states $a$ and $b$. 

Using the projection operator on the states of the spin doublet,
$P_{1} = P_{+} + P_{-}$, we obtain additional PDF:
\begin{eqnarray}
	\langle P_{0,k} P_{1,k+l} \rangle
	&=&
	\left( \frac{1}{4} - m^2 \right)
	\left( 
	1 - \Lambda_0^{l}  
	\right)
	,
	\label{eq:P01l}
	\\
	\langle P_{1,k} P_{1,k+l} \rangle
	&=&
	\left( \frac{1}{2} - m \right)^2
	+ \left( \frac{1}{4} - m^2 \right) \Lambda_0^{l} 
	\label{eq:P11l}
	.
\end{eqnarray}

From the identity $\sigma = P_{+} - P_{-}$ we find the spin-spin correlation function:
\begin{equation}
	\langle \sigma_{k} \sigma_{k+l} \rangle
	=
	\left( \frac{1}{2} - m \right)	\Lambda_\sigma^{l}  
	.
	\label{eq:ss-cf}
\end{equation}
Assuming that $\langle \sigma_{k} \sigma_{k+l} \rangle \propto e^{-l/\ell}$, we find the expression for the spin correlation length $\ell$:
\begin{equation}
	\ell^{-1} = \ln \left| \Lambda_{\sigma}^{-1} \right| = \ell_{Is}^{-1} + \tilde{\ell}^{-1}
	,\qquad
	\tilde{\ell}^{-1} = \ln \frac{ g + \frac{1}{2} }{ g - m }
	.
	\label{eq:cl}
\end{equation}
Here $\ell_{Is}$ is the spin correlation length of the pure Ising chain, 
and $\tilde{\ell}$ is the additive due to impurities. 
The emergence of $\tilde{\ell}$ decreases the spin correlation length $\ell$.

Writing the impurity correlation function in the form $K_{00}(l) \propto e^{-l/\ell_0}$, we find the impurity correlation length $\ell_0$:
\begin{equation}
	\ell_0^{-1} = \ln \left| \Lambda_{0}^{-1} \right| 
	.
	\label{eq:cl0}
\end{equation}
If $V<|J|$, the function $\Lambda_0$ changes the sign from negative at high temperatures to positive at low temperatures. 
This means that the tendency to the charge ordering of impurities at high temperatures is replaced by a tendency to a homogeneous distribution at low temperatures. The temperature, when $\Lambda_0=0$, is obtained from the equation $e^{|K|} = e^{W} + \left( e^{2W} - 1 \right)^{1/2}$.
When $V\geq|J|$, the function $\Lambda_0$ is always negative.
\label{sec:21}

%%%%%%%%%%%%%%%%%%%%%%%%%%%%%%%%%%%%%%%%%%%%%%%%%%%%%%%%%%%%%%%%%%%%%%%%%%%%%%%%%%%%%%%%%%%%%%%%%%%
\subsection{Probability distributions for the finite sequences}

PDF $\left\langle P_{a,k} P_{b,k+1} \right\rangle$ equals to the probability for the nearest neighbors 
to be in the states $a$ and $b$ simultaneously, $P(ab)$. 
Using Bayes’s formula, it can be represented as
\begin{equation}
	\left\langle P_{a,k} P_{b,k+1} \right\rangle
	\equiv P(ab) 
	= P(a) P(b|a) 
	= P(a|b) P(b)
	,
	\label{eq:ProbPairCond}
\end{equation}
where $P(\sigma)$ is the probability for $k$th site to be in the state $\sigma$, 
$P(\sigma'|\sigma)$ is the conditional probability of the state $\sigma'$ on $(k{+}1)$st site, 
given that $k$th site is in the state $\sigma$. 
The conditional probabilities for the states with $\sigma= {+}1,0,{-}1$ can be written as a non-symmetric matrix with the elements 
$\mathcal{P}_{ab} = P(a|b)$:
\begin{equation}
	\mathcal{P} 
	=
	\frac{1}{2g{+}1}
	\begin{pmatrix}
	\displaystyle  \frac{(g{-}m) e^{K}}{\cosh K}  &  
	\displaystyle  \frac{1}{2} - m  &  
	\displaystyle  \frac{(g{-}m) e^{-K}}{\cosh K}  
	\\[1em]
	\displaystyle  1 + 2m  &  
	\displaystyle  2(g {+} m)  &  
	\displaystyle  1 + 2m  
	\\[0.5em]
	\displaystyle  \frac{(g{-}m) e^{-K}}{\cosh K}  &  
	\displaystyle  \frac{1}{2} - m  &  
	\displaystyle  \frac{(g{-}m) e^{K}}{\cosh K} 
	\end{pmatrix}
	.
	\label{eq:Pmat}
\end{equation}
$\mathcal{P}$ is a left stochastic matrix and it has eigenvalues 1, $\Lambda_0$ and $\Lambda_{\sigma}$. 
The eigenvector for the eigenvalue 1 is proportional to the equilibrium distribution 
for the states of the dilute Ising chain 
$\left( 
\frac{1}{2} \left( \frac{1}{2}{-}m \right),\, \frac{1}{2}{+}m,\, \frac{1}{2} \left( \frac{1}{2}{-}m \right) 
\right)$. 
This means that the dilute Ising chain 
corresponds to the Markov chain with the transition matrix $\mathcal{P}^{T}$. 
To verify this, one can check that PDF (\ref{eq:P00l}-\ref{eq:Ppml}) 
are expressed in terms of the $l$th power of the matrix $\mathcal{P}$:
\begin{eqnarray}
	&&
	\left\langle P_{a,k} P_{b,k+l} \right\rangle
	= \nonumber \\
	&& \quad = 
	\sum_{\sigma_{1},\ldots,\sigma_{l-1}} 
	P(a|\sigma_{1}) P(\sigma_{1}|\sigma_{2}) \ldots P(\sigma_{l-1}|b) P(b)
	\nonumber \\
	&& \quad = 
	\mathcal{P}^l_{ab} P(b)
	\;=\; \mathcal{P}^l_{ba} P(a)
	.
\end{eqnarray}
This is a consequence of the Chapman-Kolmogorov equations. 
The Markov property for the pure Ising system is well-known~\cite{Malyshev1991}. 
Similarly, PDF (\ref{eq:P00l},\ref{eq:P01l},\ref{eq:P11l}) can be calculated using the $l$th power 
of the conditional probability matrix for the states 0 (impurity) and 1 (magnetic):
\begin{equation}
	\mathcal{P}' 
	=
	\frac{2}{2g+1}
	\left(
	\begin{array}{cc}
	\displaystyle g + m  &  
	\displaystyle \frac{1}{2} + m  \\
	\displaystyle \frac{1}{2} - m  &  
	\displaystyle g - m
	\end{array}
	\right)
	.
	\label{eq:PmatPrime}
\end{equation}

The Markov property allows us to calculate the probability of any given sequence of states $(ab \ldots cd)$ 
for the sites following each other in a dilute Ising chain using the matrix elements of 
$\mathcal{P}$ or $\mathcal{P}'$: 
\begin{eqnarray}
	P(ab \ldots cd)
	&=& P(a|b) \ldots P(c|d) P(d) 
	\nonumber \\ 
	&=& P(d|c) \ldots P(b|a) P(a)
	.
	\label{eq:Pabcd}
\end{eqnarray}

Now consider an isolated sequence consisting of $l$ repeating blocks $\boldsymbol{\sigma}$, 
where $\boldsymbol{\sigma} = (\sigma_1 \ldots \sigma_n)$. 
The probability that this sequence exists is
\begin{equation}
	P_l(\boldsymbol{\sigma})  
	= \sum_{ \boldsymbol{\sigma}'\neq\boldsymbol{\sigma},	\boldsymbol{\sigma}''\neq\boldsymbol{\sigma} }
	P ( \boldsymbol{\sigma}'  \boldsymbol{\sigma}^{l}  \boldsymbol{\sigma}'' )
	= P_1(\boldsymbol{\sigma}) \, q_{\boldsymbol{\sigma}}^{l-1}
	,
\end{equation}
where
\begin{equation}
	q_{\boldsymbol{\sigma}}^{}
	= P(\sigma_1|\sigma_2) \ldots P(\sigma_n|\sigma_1)
	,
	\label{eq:q}
\end{equation}
\begin{equation}
	P_1(\boldsymbol{\sigma})  
	= \sum_{ \boldsymbol{\sigma}'\neq\boldsymbol{\sigma},	\boldsymbol{\sigma}''\neq\boldsymbol{\sigma} }
	P ( \boldsymbol{\sigma}' \boldsymbol{\sigma} \boldsymbol{\sigma}'' )
	.
	\label{eq:A}
\end{equation}
Here $q_{\boldsymbol{\sigma}}^{}$ can be treated as the probability of a cyclic sequence consisting of sites 
$\sigma_1 \ldots \sigma_n$ 
and the accounting of equation (\ref{eq:Pabcd}) gives inequality 
$q_{\boldsymbol{\sigma}}^{}<P(\boldsymbol{\sigma})$. 
$P_1(\boldsymbol{\sigma})$ is the probability of an isolated block $\boldsymbol{\sigma}$, 
and calculating the sum in (\ref{eq:A}) we get 
\begin{equation}
	P_1(\boldsymbol{\sigma}) = \left( 1 - q_{\boldsymbol{\sigma}}^{} \right)^2   P(\boldsymbol{\sigma})
	.
	\label{eq:P1}
\end{equation}
Equation (\ref{eq:P1}) generalizes the result of Yilmaz and Zimmermann~\cite{Yilmaz2005} for a pure Ising chain. 
The sum of $P_l(\boldsymbol{\sigma})$ over $l$ is the total probability of the boundary configurations  
$(\boldsymbol{\sigma}' \boldsymbol{\sigma})$:
\begin{equation}
	\sum_{l=1}^{\infty} P_l(\boldsymbol{\sigma}) 
	= \sum_{ \boldsymbol{\sigma}'\neq\boldsymbol{\sigma} }
	P ( \boldsymbol{\sigma}' \boldsymbol{\sigma} )
	= \left( 1 - q_{\boldsymbol{\sigma}}^{} \right)  P(\boldsymbol{\sigma}) 
	.
	\label{eq:sumPl}
\end{equation}
Using the quantity (\ref{eq:sumPl}) as a normalization factor, 
we obtain the probability distribution for the lengths $l$ of sequences of repeating blocks $\boldsymbol{\sigma}$:
\begin{equation}
	p_l(\boldsymbol{\sigma}) = \left( 1 - q_{\boldsymbol{\sigma}}^{} \right)  q_{\boldsymbol{\sigma}}^{l-1}
	.
	\label{eq:geom_dist}
\end{equation}
This is a geometric distribution with $q_{\boldsymbol{\sigma}}^{}$ defined by (\ref{eq:q}). 
The mean length and the length dispersion of the sequences of repeating blocks $\boldsymbol{\sigma}$ are
\begin{equation}
	\bar{l}_{\boldsymbol{\sigma}} = \frac{1}{ 1 - q_{\boldsymbol{\sigma}}^{} }
	,\quad
	D(l_{\boldsymbol{\sigma}}) = \frac{ q_{\boldsymbol{\sigma}}^{} }{ \left( 1 - q_{\boldsymbol{\sigma}}^{} \right)^2 }
	.
\end{equation}

%%%%%%%%%%%%%%%%%%%%%%%%%%%%%%%%%%%%%%%%%%%%%%%%%%%%%%%%%%%%%%%%%%%%%%%%%%%%%%%%%%%%%%%%%%%%%%%%%%%
%%%%%%%%%%%%%%%%%%%%%%%%%%%%%%%%%%%%%%%%%%%%%%%%%%%%%%%%%%%%%%%%%%%%%%%%%%%%%%%%%%%%%%%%%%%%%%%%%%%
\section{Results}

%%%%%%%%%%%%%%%%%%%%%%%%%%%%%%%%%%%%%%%%%%%%%%%%%%%%%%%%%%%%%%%%%%%%%%%%%%%%%%%%%%%%%%%%%%%%%%%%%%%
\begin{figure*}
\centering
	\includegraphics[width=0.85\textwidth]{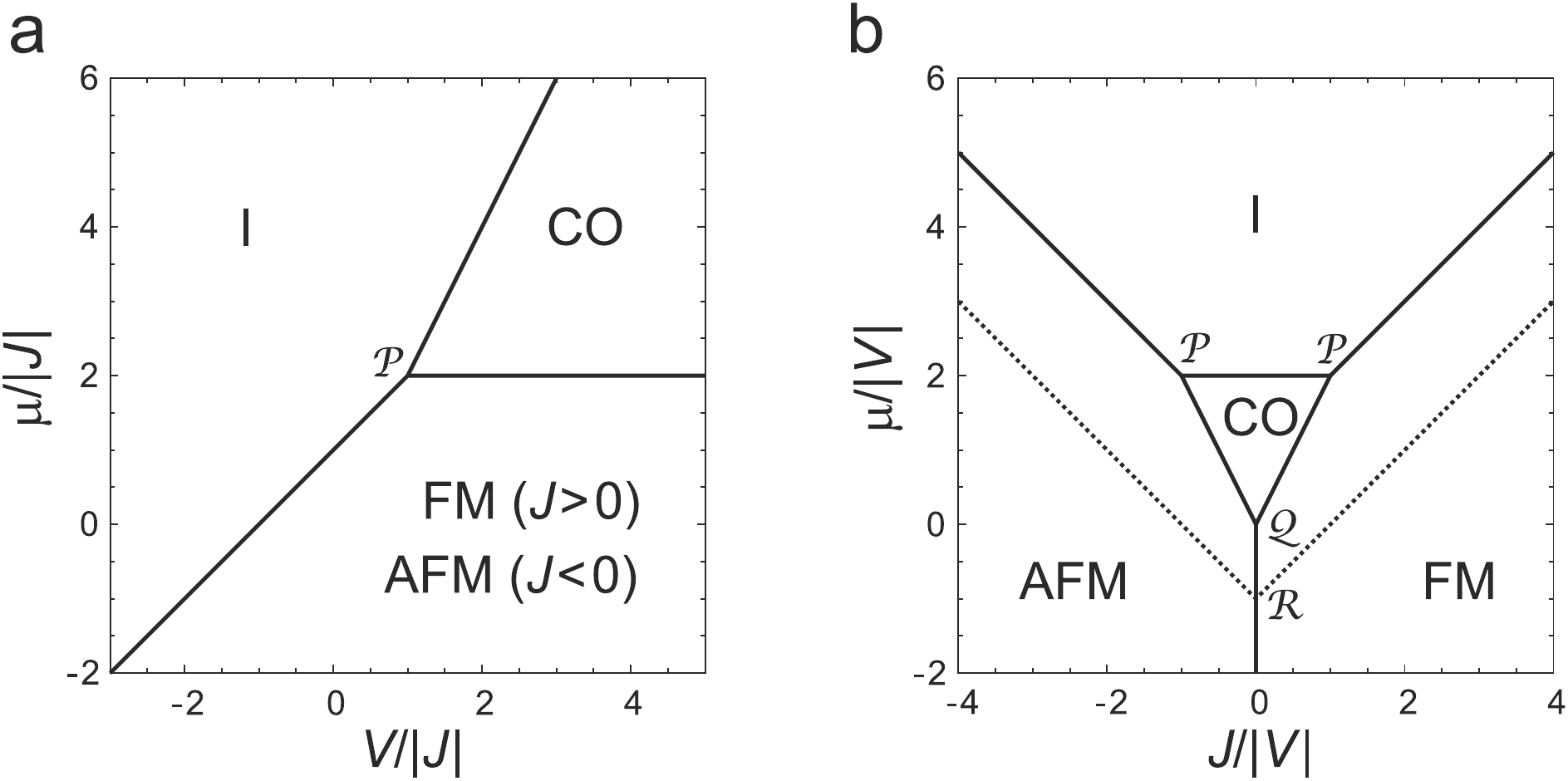}
	\caption{The phase diagrams of the 1D dilute Ising model (\ref{eq:H}) at zero temperature 
	(a) in plane ($V/|J|$, $\mu/|J|$), (b) in plane ($J/|V|$, $\mu/|V|$). 
	The impurity (I), charge-ordered (CO), ferromagnetic (FM), and antiferromagnetic (AFM) phases 
	correspond to the following configurations of the $z$-components of the neighbour pseudospins:
	I $\rightarrow$ (0,0), 
	CO $\rightarrow$ (0,$\pm1$), 
	FM $\rightarrow$ ($\pm1$,$\pm1$), 
	AFM $\rightarrow$ ($1$,$-1$). 
	(a) Solid lines are the coexistence curves of phases I, CO, and FM for $J>0$ or phases I, CO, and AFM for $J<0$. 
	The point $\mathcal{P}$($V/|J|=1$ and $\mu/|J|=2$) is a tricritical point. 
	(b) Solid lines show the coexistence curves of phases I, CO, FM, and AFM for $V>0$. 
	Along with $\mathcal{P}$, the tricritical point $\mathcal{Q}$($J/|V|=0$ and $\mu/|V|=0$) is shown. 
	Dashed lines show the coexistence curves of phases I and FM and phases I and AFM for $V<0$. 
	In the latter case, the only tricritical point is $\mathcal{R}$($J/|V|=0$ and $\mu/|V|=-1$).
	\label{fig:gspd}
	}
\end{figure*}
%%%%%%%%%%%%%%%%%%%%%%%%%%%%%%%%%%%%%%%%%%%%%%%%%%%%%%%%%%%%%%%%%%%%%%%%%%%%%%%%%%%%%%%%%%%%%%%%%%%

In Fig.~\ref{fig:gspd} we present the phase diagrams of the 1D dilute Ising model (\ref{eq:H}) at zero temperature in planes ($V/|J|$, $\mu/|J|$) and ($J/|V|$, $\mu/|V|$). 
The impurity (I), charge-ordered (CO), ferromagnetic (FM), and antiferromagnetic (AFM) phases 
correspond to the following configurations of the $z$-components of the neighbour pseudospins:
I $\rightarrow$ (0,0), 
CO $\rightarrow$ (0,1) or (0,$-1$), 
FM $\rightarrow$ (1,1) or ($-1$,$-1$), 
AFM $\rightarrow$ (1,$-1$). 
The values of the grand potential per one site at zero temperature for each phase are given by the expressions:
$\omega_{I} = V - \mu$,
$\omega_{CO} = - \mu/2$, 
$\omega_{FM} = - J$, 
$\omega_{AFM} = J$. 
This defines the concentration of impurities $n$ in each phase: 
$n_{I} = 1$,
$n_{CO} = 1/2$, 
$n_{FM} = 0$, 
$n_{AFM} = 0$. 
Intermediate values of $n$ correspond to the states with phase separation, which are represented by the coexistence curves
FM/CO and AFM/CO ($0<n<1/2$), CO/I ($1/2<n<1$), FM/I and AFM/I ($0<n<1$). 
The tricritical point $\mathcal{P}$ corresponds to the mixture of phases I, CO, and FM at $J>0$ (or I, CO, and AFM at $J<0$) 
when $V>0$ and $V=|J|$. 
The frustrated states for all $n$, $0<n<1$, arise because of the equal strength of the spin-spin and impurity-impurity interactions. 
The tricritical points $\mathcal{Q}$ and $\mathcal{R}$ represent completely disordered magnetic states since they correspond to the value $J = 0$.

\begin{figure}
\centering
	\includegraphics[width=0.45\textwidth]{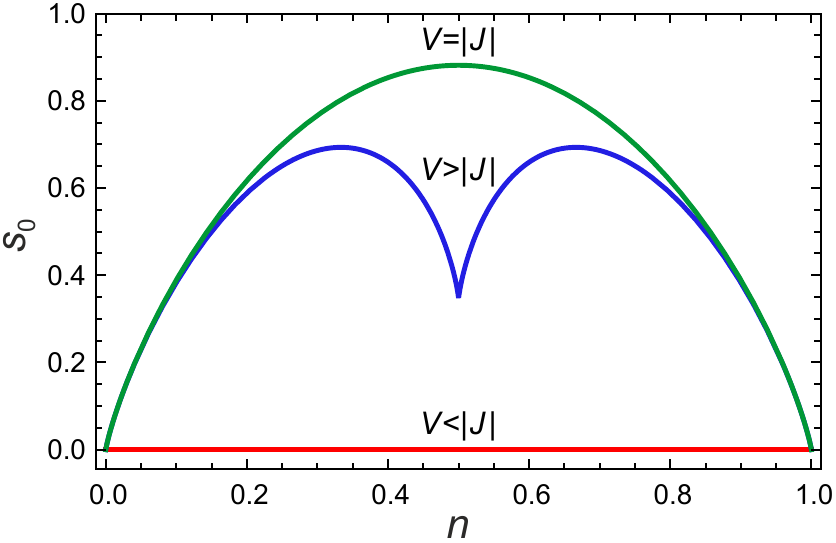}
	\caption{(color online)  An entropy of the 1D dilute Ising model at zero temperature for $V<|J|$, $V=|J|$ and $V>|J|$.
	\label{fig:s0}
	}
\end{figure}

The qualitative difference of states at zero temperature is well illustrated by the entropy of the dilute Ising chain, defined as a function of temperature and impurities concentration by the following expression
\begin{eqnarray}
	s &=& 
	{}- m \ln \left( 2 \, \frac{ g + m  }{  g - m  }  \, e^{W} \cosh K  \right) 
	- {} 
	\nonumber \\
	&&
	{}- \frac{1}{2}  \ln \frac{ \frac{1}{4} - m^2 }{ 2 \left( g + \frac{1}{2} \right)^2 }
	+ \left( m + \frac{1}{2} \right) \frac{g + m}{g + \frac{1}{2}} \, W 
	+{}
	\nonumber\\
	&&
	+ \left( m - \frac{1}{2} \right) \frac{g - m}{g + \frac{1}{2}} \, K \tanh K
	.
	\label{eq:s}
\end{eqnarray}
The limiting values $s_0$ of entropy at $\theta=0$ are listed in Table~\ref{tab:s0} and 
are shown as a dependencies of the impurities concentration in Fig.\ref{fig:s0}. 
We see that $s_0$ depends only on the concentration of impurities 
in different ways for $V<|J|$, $V=|J|$ and $V>|J|$, 
but does not depend on the interaction parameters themselves. 
If $V<|J|$, we get $s_0 = 0$, but if $V\geq|J|$, the entropy of the ground state is greater than zero, so
these states should be referred to in modern language as frustrated \cite{Zarubin2019}. 
When $V=|J|$, the maximum value of entropy is $s_0 = \ln\left( 1 + \sqrt{2} \right) \approx 0.881$ for $m=0$, 
while in the case $V>|J|$, the entropy has two maxima $s_0 = \ln 2 \approx 0.693$ at $|m|=\frac{1}{6}$
and the local minimum $s_0 = \frac{1}{2}\ln 2 \approx 0.347$ at $m=0$.

%%%%%%%%%%%%%%%%%%%%%%%%%%%%%%%%%%%%%%%%%%%%%%%%%%%%%%%%%%%%%%%%%%%%%%%%%%%%%%%%%%%%%%%%%%%%%%%%%%%
%%%%%%%%%%%%%%%%%%%%%%%%%%%%%%%%%%%%%%%%%%%%%%%%%%%%%%%%%%%%%%%%%%%%%%%%%%%%%%%%%%%%%%%%%%%%%%%%%%%

%%%%%%%%%%%%%%%%%%%%%%%%%%%%%%%%%%%%%%%%%%%%%%%%%%%%%%%%%%%%%%%%%%%%%%%%%%%%%%%%%%%%%%%%%%%%%%%%%%%
{
\renewcommand{\tabcolsep}{0.4em}

\begin{table}%[htbp]
\centering
		\begin{tabular}{lc}
			\hline\hline
			\rule{0em}{1.5em}%
			Parameters  &  $ s_0 $  
			\\[0.5em]
		\hline
			$ V<|J| $ \rule{0em}{2em}  &  $ 0 $ 
			\\[1.5em]
			$ V=|J| $  
			&  $ \displaystyle  
			\ln  \left(  \frac{1}{2} + g_0  \right) 
			+ \frac{1}{2}  \ln \frac{2}{ \frac{1}{4} - m^2 }
			- m  \ln  \frac{ g_0 + m }{ g_0 - m }  
			$ 
			\\[1.5em]
			$ V>|J| $
			&  $ \displaystyle 
			 \frac{1}{2}   \ln \frac{ 1 + 2|m| }{ 1 - 2|m| } 
			 +  |m|  \ln \frac{ \frac{1}{4} - m^2 }{ 8 m^2 } 
			 + \frac{1}{2}  \ln 2  
			$ 
			\\[1.5em]
			\hline\hline
		\end{tabular}
		\caption{
		\label{tab:s0}
		The zero-temperature value $s_0$ of entropy defined by equation (\ref{eq:s}) 
		for different sets of parameters of the dilute Ising chain. 
		}
\end{table}

}
%%%%%%%%%%%%%%%%%%%%%%%%%%%%%%%%%%%%%%%%%%%%%%%%%%%%%%%%%%%%%%%%%%%%%%%%%%%%%%%%%%%%%%%%%%%%%%%%%%%

%%%%%%%%%%%%%%%%%%%%%%%%%%%%%%%%%%%%%%%%%%%%%%%%%%%%%%%%%%%%%%%%%%%%%%%%%%%%%%%%%%%%%%%%%%%%%%%%%%%
{
\renewcommand{\tabcolsep}{0.8em}

\begin{table}%[thbp]
\centering
		\begin{tabular}{lcc}
			\hline\hline
			\rule{0em}{1.5em}%
			Parameters
		  & $ \displaystyle  \ell $ 
			& $ \displaystyle  \ell_0  $ 
			\\[0.5em]
			\hline
			\rule{0em}{2em}%
			\begin{tabular}[c]{l}  $V<|J|$  \end{tabular}  
		  &  $ \displaystyle  \propto  e^{ \frac{|K|-W}{2} } $ 
			&  $ \displaystyle  \propto  e^{ \frac{|K|-W}{2} } $ 
			\\[1.5em]
			\begin{tabular}[c]{l}  $V=|J|$  \end{tabular}  
		  &  $ \displaystyle  \left( \ln \frac{g_0 + \frac{1}{2}}{g_0 - m} \right)^{\!\!-1} $ 
			&  $ \displaystyle  \left(  \ln \frac{1 + 2g_0}{1 - 2g_0} \right)^{\!\!-1} $ 
			\\[2em]
			\begin{tabular}[c]{l}  $V>|J|$ \\ $m<0$ \end{tabular}  
		  &  $ \displaystyle  \left( \ln \frac{|m| + \frac{1}{2}}{2|m|} \right)^{\!\!-1} $ 
			&  $ \displaystyle  \left(  \ln \frac{1 + 2|m|}{1 - 2|m|}  \right)^{\!\!-1} $ 
			\\[2em]
			\begin{tabular}[c]{l}  $V>|J|$ \\ $m=0$ \end{tabular}  
		  &  $ 0 $ 
			&  $ \displaystyle  \propto  e^{ \frac{W-|K|}{2} }  $ 
			\\[1.5em]
			\begin{tabular}[c]{l}  $V>|J|$ \\ $m>0$ \end{tabular}  
		  &  $ 0 $ 
			&  $ \displaystyle  \left(  \ln \frac{1 + 2|m|}{1 - 2|m|}  \right)^{\!\!-1} $ 
			\\[1.5em]
			\hline\hline
		\end{tabular}
		\caption{
		\label{tab:ll0}
		Asymptotic behavior of the spin correlation length $\ell$ 
		and the impurity correlation length $\ell_0$ 
		at zero temperature for different sets of parameters. 
		}
\end{table}

}
%%%%%%%%%%%%%%%%%%%%%%%%%%%%%%%%%%%%%%%%%%%%%%%%%%%%%%%%%%%%%%%%%%%%%%%%%%%%%%%%%%%%%%%%%%%%%%%%%%%

\begin{figure*}
\centering
	\includegraphics[width=0.85\textwidth]{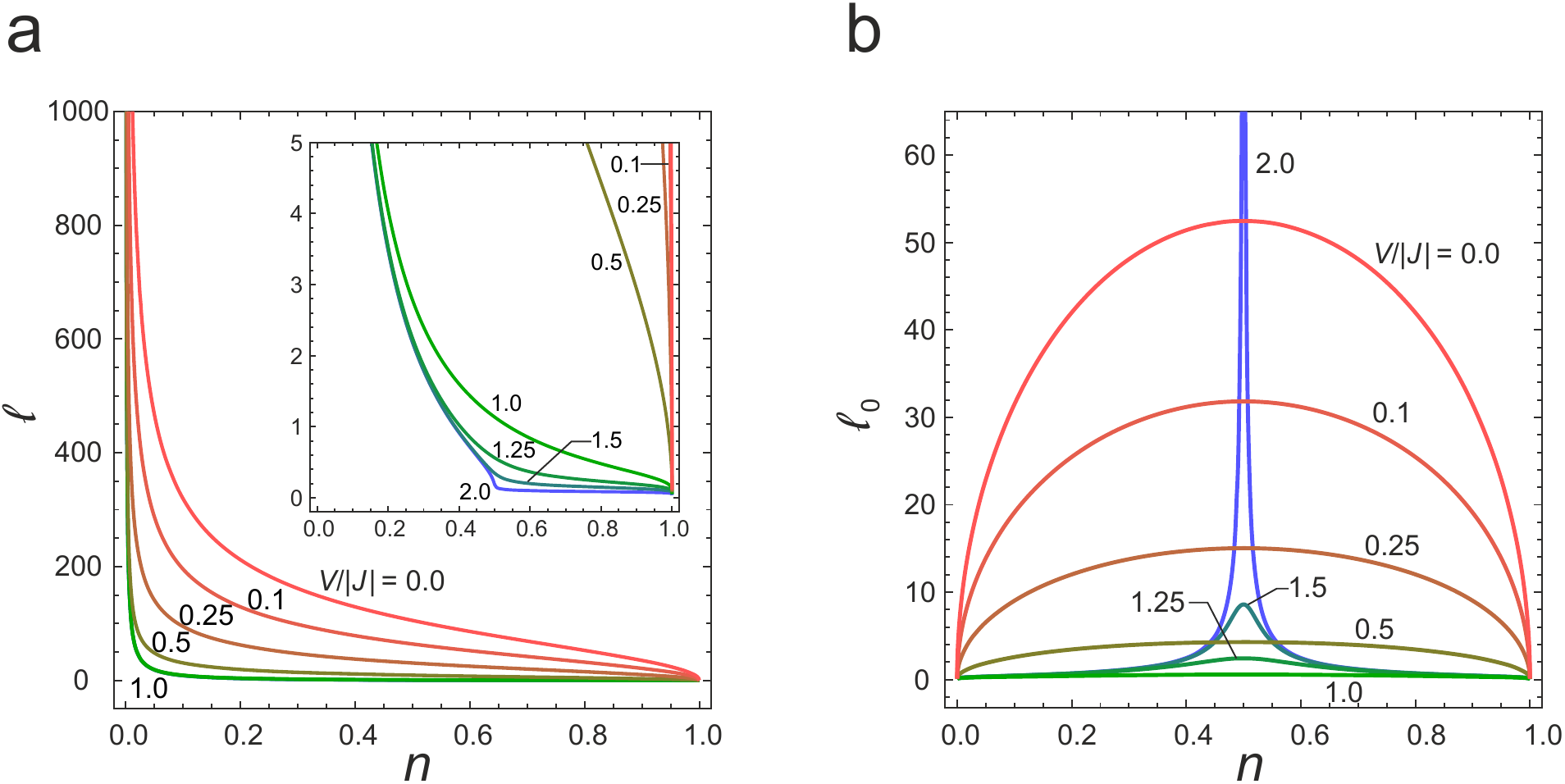}
	\caption{(color online) Concentration dependences of 
	(a) the spin correlation length $\ell$ and 
	(b) the impurity correlation length $\ell_0$
	at $\theta/|J|=0.1$ for different values of $V/|J|$.
	The insert in (a) shows that the spin correlation length tends to zero at low temperatures for $V>|J|$ and $n\geq0.5$.
	\label{fig:ll0}
	}
\end{figure*}

Concentration dependences of the spin correlation length and the impurity correlation length 
at low temperature ($\theta/|J|=0.1$)
are shown in Fig.\ref{fig:ll0} for certain values of $V/|J|$. 
The analytical expressions of the low-temperature asymptotics of $\ell$ and $\ell_0$ 
are presented in Table~\ref{tab:ll0}. 
The parameter $g_0$ is given by
\begin{equation}
	g_0 = \frac{1}{\sqrt{2}} \left( \frac{1}{4} + m^2 \right)^{1/2} 
	.
	\label{eq:g0}
\end{equation}
Like other thermodynamic properties of the dilute Ising chain, 
the low-temperature limits of $\ell$ and $\ell_0$ are qualitatively different for $V<|J|$, $V=|J|$, and $V>|J|$.
If $V<|J|$, both $\ell$, and $\ell_0$ tend to infinity with the same asymptotic behavior.
This means that with decreasing temperature the system is divided into macroscopically homogeneous domains consisting of non-magnetic impurities and magnetic sites, respectively.
In this case, the magnetic matrix pushes impurities, thereby minimizing the surface energy. 
Another quasi-ordered situation appears at $V>|J|$ and $m=0$ ($n=1/2$), when the impurity correlation length tends to infinity in contrast to the spin correlation length which tends to zero.
This indicates the formation of charge ordering at zero temperature when the sites occupied by non-magnetic charged impurities with $\sigma=0$ alternate with the sites occupied by the spin states with $\sigma=\pm1$. 
Due to the short-range character of the exchange interaction, the spin subsystem becomes an ideal paramagnet.

%%%%%%%%%%%%%%%%%%%%%%%%%%%%%%%%%%%%%%%%%%%%%%%%%%%%%%%%%%%%%%%%%%%%%%%%%%%%%%%%%%%%%%%%%%%%%%%%%%%
%%%%%%%%%%%%%%%%%%%%%%%%%%%%%%%%%%%%%%%%%%%%%%%%%%%%%%%%%%%%%%%%%%%%%%%%%%%%%%%%%%%%%%%%%%%%%%%%%%%

An analysis of the correlation lengths makes it interesting to study 
sequences of ferromagnetically ordered spins with $\boldsymbol{\sigma}=(\pm)$, 
antiferromagnetic sequences with $\boldsymbol{\sigma}=({+}{-})$,
sequences of impurities with $\boldsymbol{\sigma}=(0)$, 
and charge-ordered sequences with $\boldsymbol{\sigma}=(01)$. 
We give the expressions of the parameter $q_{\boldsymbol{\sigma}}$ 
in the geometric distribution (\ref{eq:geom_dist}) for these sequences in Table~\ref{tab:Q}.

%%%%%%%%%%%%%%%%%%%%%%%%%%%%%%%%%%%%%%%%%%%%%%%%%%%%%%%%%%%%%%%%%%%%%%%%%%%%%%%%%%%%%%%%%%%%%%%%%%%
{
\renewcommand{\tabcolsep}{0.5em}

\begin{table}
\centering
	\begin{tabular}{lcccc}
	\hline\hline
	\rule{0em}{1.5em}%
	$\boldsymbol{\sigma}$ \rule{0em}{1.5em} & $(0)$ & $({\pm})$ & $({+}{-})$ & $(01)$ 
	\\[1em]\hline
	\rule{0em}{2.5em}%
	$q_{\boldsymbol{\sigma}}^{}$
	&  $ \displaystyle  \frac{g{+}m}{g{+}\frac{1}{2}} $  
	&  $ \displaystyle  \frac{ (g{-}m) e^{K} }{ \left(2g{+}1\right) \cosh K } $ 
	&  $ \displaystyle  \frac{ (g{-}m)^2 e^{-2K} }{ \left(2g{+}1\right)^{\!2} \! \cosh^2 \! K } $
	&  $ \displaystyle  \frac{\frac{1}{4}{-}m^2}{\left(g{+}\frac{1}{2}\right)^{\!2}} $ 
	\\[1.5em]
	\hline\hline
	\end{tabular}
	\caption{
	\label{tab:Q}
		The parameter $q_{\boldsymbol{\sigma}}^{}$ 
		of geometric distribution (\ref{eq:geom_dist}) 
		for sequences of repeating blocks $\boldsymbol{\sigma}$. 
	}
\end{table}

}
%%%%%%%%%%%%%%%%%%%%%%%%%%%%%%%%%%%%%%%%%%%%%%%%%%%%%%%%%%%%%%%%%%%%%%%%%%%%%%%%%%%%%%%%%%%%%%%%%%%

The low-temperature properties of the mean lengths of repeating sequences are listed in Table~\ref{tab:meanL0}. 

If $V<|J|$, the mean length of the impurity sequences $\bar{l}_{(0)}$ diverges at $\theta\to0$,
having the same asymptotic behavior as the impurity correlation length $\ell_0$.
If $V\geq|J|$, $\bar{l}_{(0)}$ remains finite at zero temperature, 
and if $m \leq 0$, the mean length approaches its minimum value, 
$\bar{l}_{(0)}=1$, that corresponds to single impurities separated by the chain sites with $\sigma=\pm1$.

%%%%%%%%%%%%%%%%%%%%%%%%%%%%%%%%%%%%%%%%%%%%%%%%%%%%%%%%%%%%%%%%%%%%%%%%%%%%%%%%%%%%%%%%%%%%%%%%%%%
{
\renewcommand{\tabcolsep}{0.7em}

\begin{table}%[htbp]
\centering
		\begin{tabular}{lccc}
			\hline\hline
			\rule{0em}{1.5em}%
			Parameters
		  &  $ \displaystyle  \bar{l}_{(0)} $ 
			&  $ \displaystyle  \bar{l}_{(\pm)} $ 
			&  $ \displaystyle  \bar{l}_{(01)} $ 
			\\[0.75em]
		\hline
			\begin{tabular}[c]{l}  $V<|J|$  \end{tabular}  
			\rule{0em}{2em} 
		  &  $ \displaystyle  \propto e^{\frac{|K| - W}{2}} $ 
			&  $ \displaystyle  \propto e^{\frac{|K| - W}{2}} $
			&  $  1 $ 
			\\[1.5em]
			\begin{tabular}[c]{l}  $V=|J|$  \end{tabular}  
		  &  $ \displaystyle  \frac{2g_0+1}{1-2m} $ 
			&  $ \displaystyle  \frac{2g_0+1}{1+2m} $
			&  $ \displaystyle  \frac{ g_0^2+g_0+\frac{1}{4} }{ g_0^2+g_0+m^2 } $ 
			\\[2em]
			\begin{tabular}[c]{l}  $V>|J|$ \\ $m \neq 0$ \end{tabular}  
		  &  $ \displaystyle  \frac{ 1 + 2|m| }{1-2m} $ 
			&  $ \displaystyle  \frac{ 1 + 2|m| }{1+2m} $
			&  $ \displaystyle  \frac{\frac{1}{2} + |m|}{2|m|}  $ 
			\\[2em]
			\begin{tabular}[c]{l}  $V>|J|$ \\ $m = 0$ \end{tabular}  
		  &  $  1 $ 
			&  $  1 $ 
			&  $ \displaystyle  \propto e^{\frac{W-|K|}{2}} $ 
			\\[1.5em]
			\hline\hline
		\end{tabular}
		\caption{
		\label{tab:meanL0}
		Asymptotic behavior of the mean lengths 
		for the impurity sequences $(0)^l$, 
		ferromagnetic sequences $(\pm)^l$ 
		and the charge-ordered sequences $(01)^l$ 
		at zero temperature. 
		}
\end{table}

}
%%%%%%%%%%%%%%%%%%%%%%%%%%%%%%%%%%%%%%%%%%%%%%%%%%%%%%%%%%%%%%%%%%%%%%%%%%%%%%%%%%%%%%%%%%%%%%%%%%%

Asymptotic behavior of $\bar{l}_{(\pm)}$ and $\bar{l}_{({+}{-})}$ at $\theta\to0$ depends on the sign of $J$. 
If $J<0$ ($J>0$), we obtain $\bar{l}_{(\pm)}=1$ ($\bar{l}_{({+}{-})}=1$) at $\theta=0$ for any parameters of the dilute Ising chain.
If $J>0$, asymptotic expressions for ferromagnetic sequences and for the impurity sequences 
differ by replacing $m$ with $-m$. 
The value of $\bar{l}_{(\pm)}$ diverges for $V<|J|$ and has the same asymptotic behavior as the spin correlation length $\ell$. 
If $V\geq|J|$, $\bar{l}_{(\pm)}$ remains finite at zero temperature, 
and if $m \geq 0$, we obtain $\bar{l}_{(\pm)}=1$. 
If $J<0$, asymptotic behavior of $\bar{l}_{({+}{-})}$ 
is the same as of $\bar{l}_{(\pm)}$ for $J>0$ 
with some difference in amplitude due to the number of sites in the minimum block of ferromagnetic and antiferromagnetic sequences.

The low-temperature limit of $\bar{l}_{(01)}$ at $V<|J|$ 
complements the previous results for the impurity and spin sequences 
since the pair $(01)$ is just the boundary between them. 
The concentration of pairs $(01)$
\begin{equation}
	\left\langle P_{0,k} P_{1,k+1} \right\rangle = P(01)
	= \frac{ \frac{1}{2} - 2 m^2 }{ 2g + 1 }
\end{equation}
becomes zero only at $\theta=0$, 
which corresponds to the infinity length of the impurity and the spin sequences. 
If $V=|J|$ or $V>|J|$, $m\neq0$, $\bar{l}_{(01)}$ is finite, 
and only if $V>|J|$ and $m=0$, 
the mean length for the charge ordered sequences tends to infinity at $\theta\to0$. 
The asymptotic behavior of $\bar{l}_{(01)}$ determines the asymptotic behavior of the impurity correlation length $\ell_0$ in this case.

The results found in this section specify details of the ordering processes in the dilute Ising chain 
with decreasing temperature. 
For repeating sequences $\boldsymbol{\sigma}^l$, 
the geometric distribution $p_l$ of lengths $l$ realizes, 
$p_l = (1-q_{\boldsymbol{\sigma}}) q_{\boldsymbol{\sigma}}^{l-1}$. 
The mean length $\bar{l}_{\boldsymbol{\sigma}} = (1-q_{\boldsymbol{\sigma}})^{-1}$ 
tends to infinity only if $q_{\boldsymbol{\sigma}}$ goes to unity, 
and the dispersion of the cluster lengths also tends to infinity, asymptotically as a square of 
$\bar{l}_{\boldsymbol{\sigma}}$: 
$D(l_{\boldsymbol{\sigma}}) 
= q_{\boldsymbol{\sigma}} (1-q_{\boldsymbol{\sigma}})^{-2} \propto \bar{l}_{\boldsymbol{\sigma}}^2$. 
The final ordering at zero temperature is achieved only when the concentration of the boundaries of the sequences becomes zero.
This is similar to the well-known result of the percolation theory, 
where the site percolation threshold in the 1D problem equals unity.

\begin{figure*}
\centering
\includegraphics[width=0.75\textwidth]{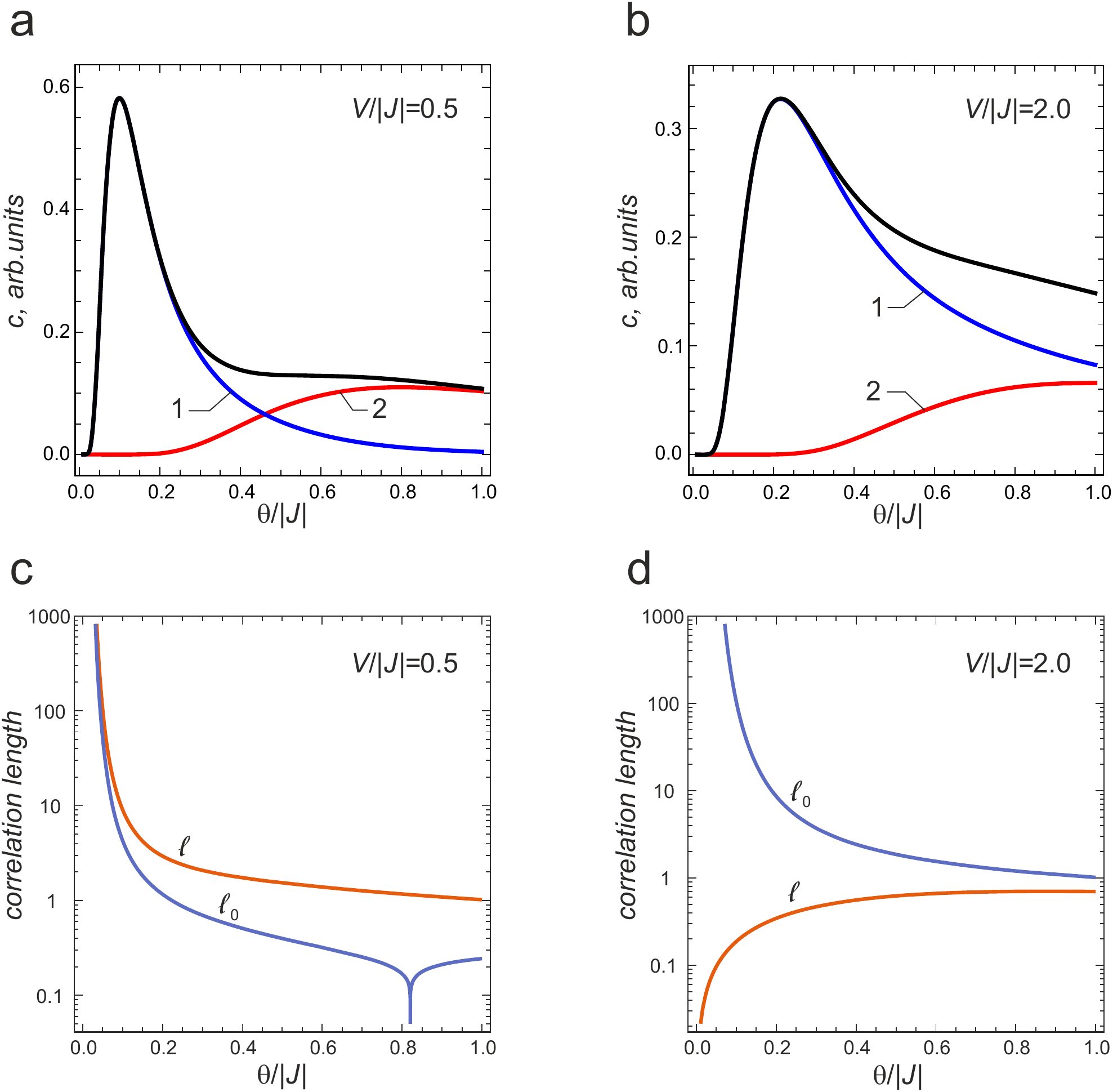}
\caption{
(color online) 
Specific heat and the contributions of first {(--~1)} and second {(--~2)} terms in equation (\ref{eq:c12}) 
for $m=0$ ($n=1/2$) (a) at $V/|J|=0.5$ and (b) at $V/|J|=2.0$. 
The spin correlation length $\ell$ and the impurity correlation length $\ell_0$ 
for $m=0$ (c) at $V/|J|=0.5$ and (d) at $V/|J|=2.0$. 
The $y$-axis in (c) and (d) has a logarithmic scale. 
The peculiarity of $\ell_0$ near $\theta/|J|=0.8$ in (c) corresponds to the solution of the equation $\Lambda_0 = 0$ 
discussed at the end of Section \ref{sec:21}.
\label{fig:c12}  
}
\end{figure*}

It is worth noting that
both at $V<|J|$ and at $V>|J|$, 
the specific heat of the dilute Ising chain has the low-temperature peak that can be directly related to the concentration of the impurity–spin pairs, 
$\left\langle P_{0,k} P_{1,k+1} \right\rangle$. 
The internal energy can be written using PDF in the form:
\begin{equation}
	\varepsilon = 
	- J \left\langle  P_{1,k} P_{1,k+1} \right\rangle \tanh K + V \left\langle  P_{0,k} P_{0,k+1} \right\rangle
	,
\end{equation}
where we use the relation 
$\left\langle \sigma_{k} \sigma_{k+1} \right\rangle = \left\langle P_{1,k} P_{1,k+1} \right\rangle \tanh K$. 
Differentiating by temperature the identity 
$	\left\langle  P_{0,k} P_{0,k+1} \right\rangle + \left\langle  P_{1,k} P_{1,k+1} \right\rangle + 2 \left\langle  P_{0,k} P_{1,k+1} \right\rangle = 1$
we find
\begin{equation}
	\left\langle  P_{0,k} P_{0,k+1} \right\rangle_{\theta}'
	= \left\langle  P_{1,k} P_{1,k+1} \right\rangle_{\theta}'
	= - \left\langle  P_{0,k} P_{1,k+1} \right\rangle_{\theta}'
	.
	\label{eq:dPP}
\end{equation}
The first equality in (\ref{eq:dPP}) can be obtained from equations (\ref{eq:P00l}) and (\ref{eq:P11l}).
Hence the specific heat takes the form
\begin{eqnarray}
	c 
	&=& 
	\left\langle  P_{0,k} P_{1,k+1} \right\rangle_{\theta}'  \left( J \tanh K - V \right) 
	+ {}
	\nonumber \\
	&&
	{} + \left\langle  P_{1,k} P_{1,k+1} \right\rangle  \frac{K^2}{\cosh^2 K}
	.
	\label{eq:c12}
\end{eqnarray}
The contribution to the specific heat of the first and second terms in (\ref{eq:c12}) 
is shown in Fig.\ref{fig:c12}a,b. 
The expression for the specific heat of the pure Ising chain arises from the second term. 
We see, that the low-temperature peak for $n\neq0$ is mainly determined 
by the maximum rate of change with the temperature of the concentration of the impurity–spin pairs.
Using the low-temperature asymptotics, 
we can obtain a simple estimation for the specific heat maximum due to the impurities ordering 
at $V<|J|$ and at $V>|J|$, $m=0$:
\begin{equation}
	\theta_0 = \frac{  \left| |J| - V \right|  }{4} 
	.
\end{equation}

As is known, the correlation length determines the characteristic spatial scale of the system near the critical point. 
Indeed, at $\theta = 0$, the asymptotics of the spin correlation length $\ell$ coincides 
with the asymptotics of $\bar{l}_{(\pm)}$ at $J>0$ or $\bar{l}_{({+}{-})}$ at $J<0$ for $V<|J|$.  
Also the asymptotics of the impurity correlation length $\ell_0$ coincides 
with the asymptotics of $\bar{l}_{(0)}$ for $V<|J|$ or $\bar{l}_{(01)}$ for $V>|J|$. 
But, as shown in Fig.\ref{fig:c12}c,d, 
the peak of the specific heat corresponds to small values of the correlation lengths 
and is not associated with any features of the temperature dependences of $\ell$ and $\ell_0$.
However, this peak can be associated with the mean length of the impurity sequences $\bar{l}_{(0)}$, 
as $1/\bar{l}_{(0)}$ is proportional to the concentration of the impurity–spin pairs:
\begin{equation}
	\frac{1}{\bar{l}_{(0)}} 
	= \frac{2}{1 + 2m}  \left\langle P_{0,k} P_{1,k+1} \right\rangle
	.
\end{equation}

%%%%%%%%%%%%%%%%%%%%%%%%%%%%%%%%%%%%%%%%%%%%%%%%%%%%%%%%%%%%%%%%%%%%%%%%%%%%%%%%%%%%%%%%%%%%%%%%%%%
%%%%%%%%%%%%%%%%%%%%%%%%%%%%%%%%%%%%%%%%%%%%%%%%%%%%%%%%%%%%%%%%%%%%%%%%%%%%%%%%%%%%%%%%%%%%%%%%%%%

It is also worth to note 
that there are no other repeating sequences $\boldsymbol{\sigma}^l$, except for previously discussed, 
for which the mean length $\bar{l}_{\boldsymbol{\sigma}}$ tends to infinity at zero temperature. 
This is a consequence of properties of the conditional probability functions defined by matrices (\ref{eq:Pmat}) and (\ref{eq:PmatPrime}). 
Their limiting values at zero temperature are given in Table~\ref{tab:Pmatrices0}. 
The parameter $q_{\boldsymbol{\sigma}}$ in the geometric distribution $p_l$ 
is the probability of the cycle $\boldsymbol{\sigma}$, 
which is defined by the finite product of conditional probabilities (\ref{eq:q}). 
As can be seen from Table~\ref{tab:Pmatrices0},
the cycles different from the impurities, 
from the spins, ordered ferromagnetically at $J>0$, 
from the spins, ordered anti-ferromagnetically at $J<0$, 
or from the impurity–spin pairs, 
have the value of $q_{\boldsymbol{\sigma}}$ less than unity at $\theta=0$, 
which gives a finite value of $\bar{l}_{\boldsymbol{\sigma}} = (1-q_{\boldsymbol{\sigma}})^{-1}$ 
at zero temperature.

In conclusion, we list briefly the main properties of Markov chains generated by the matrices in Table~\ref{tab:Pmatrices0}. 
If $V<|J|$, the Markov chains are reducible. 
For $K>0$, three classes correspond to states with $\sigma = {+}1, \, 0, \, {-}1$. 
For $K<0$, the states with $\sigma = \pm 1$ form a periodic class that indicates antiferromagnetic ordering. 
In a Markov chain with a transition matrix $\mathcal{P}'^{T}$ the states with $\sigma = \pm 1$ are considered as one magnetic state, therefore, this Markov chain has 2 classes consisting of the impurity and magnetic states. 
If $V=|J|$, all Markov chains are irreducible and aperiodic, i.e., regular. 
If $V>|J|$ and $m \neq 0$, the Markov chains are also irreducible and aperiodic, 
but when $m=0$, the Markov chains becomes periodic. 
Two classes of states, one of which consists of an impurity state and the other consists of magnetic states, have a period 2. 
In all cases, the considered Markov chains do not have transient states, but, obviously, in an external longitudinal magnetic field, one magnetic state will become transient.

As a result, the non-ordered dilute Ising chain is described by the regular Markov chain, while the ordering generates the irregular Markov chain. 
The phase separation at $V<|J|$ generates the reducible Markov chain, 
and the charge ordering at $V>|J|$, $m=0$ generates the periodic one, 
which reflects the qualitative difference between these cases.

%%%%%%%%%%%%%%%%%%%%%%%%%%%%%%%%%%%%%%%%%%%%%%%%%%%%%%%%%%%%%%%%%%%%%%%%%%%%%%%%%%%%%%%%%%%%%%%%%%%
{
\renewcommand{\tabcolsep}{0.2em}
\renewcommand{\arraycolsep}{0.15em}

\begin{table}%[phtb]
\centering
		\begin{tabular}{p{3em}ccc}
			\hline\hline
			\rule{0em}{1.25em}%
			&  \multicolumn{2}{c}{$ \mathcal{P}$} 
			&  $ \mathcal{P}' $
			\\
			\!\!Parameters &  $ K>0 $ &  $ K<0 $ &  
			\\[0.5em]
			\hline
			\rule{0em}{3em}%
			\!\!\begin{tabular}[c]{l}  $V{<}|J|$  \end{tabular}  
		  &  
			$ 
			\begin{pmatrix}
			1 & 0 & 0 \\
			0 & ~1~ & 0 \\
			0 & 0 & 1 
			\end{pmatrix} 
			$ 
			&  
			$ 
			\begin{pmatrix}
			0 & 0 & 1 \\
			0 & ~1~ & 0 \\
			1 & 0 & 0 
			\end{pmatrix} 
			$ 
			&  
			$ 
			\begin{pmatrix}
			1 & ~0 \\
			0 & ~1
			\end{pmatrix} 
			$ 
			\\[2.5em]
			\!\!\begin{tabular}[c]{l}  $V{=}|J|$  \end{tabular}  
			&  
			$   
			\begin{pmatrix}
			1{-}a & \frac{b}{2} & 0 \\
			a & 1{-}b & a   \\
			0 & \frac{b}{2} & 1{-}a
			\end{pmatrix} 
			$ 
			&  
			$    
			\begin{pmatrix}
			0 & \frac{b}{2} & 1{-}a \\
			a & 1{-}b & a   \\
			1{-}a & \frac{b}{2} & 0
			\end{pmatrix}
			$
			&  
			$   
			\begin{pmatrix}
			1{-}b & a \\
			b & 1{-}a
			\end{pmatrix} 
			$ 
			\\[2.5em]
			\!\!\begin{tabular}[c]{l}  $V{>}|J|$ \\ $m {<} 0$  \end{tabular}  
			&  
			$    
			\begin{pmatrix}
			1{-}c & \frac{1}{2} & 0 \\
			c   &  ~0~  & c \\
			0   & \frac{1}{2} & 1{-}c 
			\end{pmatrix} 
			$ 
			&  
			$    
			\begin{pmatrix}
			0   & \frac{1}{2} & 1{-}c \\
			c   &  ~0~  & c   \\
			1{-}c & \frac{1}{2} & 0 
			\end{pmatrix} 
			$ 
			&  
			$  
			\begin{pmatrix}
			0~ & c \\
			1~ & 1{-}c
			\end{pmatrix} 
			$ 
			\\[2.5em]
			\!\!\begin{tabular}[c]{l}  $V{>}|J|$ \\ $m {=} 0$  \end{tabular}  
			&  
			$    
			\begin{pmatrix}
			 0 & \frac{1}{2} & 0 \\
			 1 & ~0~ & 1 \\
			 0 & \frac{1}{2} & 0 
			\end{pmatrix} 
			$ 
			&  
			$    
			\begin{pmatrix}
			 0 & \frac{1}{2} & 0 \\ 
			 1 & ~0~ & 1 \\ 
			 0 & \frac{1}{2} & 0 
			\end{pmatrix} 
			$ 
			&  
			$ 
			\begin{pmatrix}
			0 & ~1 \\ 
			1 & ~0
			\end{pmatrix} 
			$ 
			\\[2.5em]
			\!\!\begin{tabular}[c]{l}  $V{>}|J|$ \\ $m {>} 0$  \end{tabular}  
			&  
			$    
			\begin{pmatrix}
			0 & \frac{c}{2} & 0 \\ 
			1 & ~1{-}c~ & 1 \\ 
			0 & \frac{c}{2} & 0
			\end{pmatrix} 
			$ 
			&  
			$ 
			\begin{pmatrix}
			0 & \frac{c}{2} & 0 \\ 
			1 & ~1{-}c~ & 1 \\ 
			0 & \frac{c}{2} & 0 
			\end{pmatrix} 
			$ 
			&  
			$ 
			\begin{pmatrix}
			1{-}c & ~1 \\ 
			c   & ~0 
			\end{pmatrix} 
			$ 
			\\[2em]
			\hline\hline
		\end{tabular}
		\caption{ 
		\label{tab:Pmatrices0}
		The zero temperature values of the conditional probability matrices $\mathcal{P}$ and $\mathcal{P}'$. 
		Here 
		$a = \left( 1+2m \right) / \left( 1+2g_0 \right)$, 
		$b = \left( 1-2m \right) / \left( 1+2g_0 \right)$,
		$c = \left( 1-2|m| \right) / \left( 1+2|m| \right)$, and 
		the expression of $g_0$ is defined by equation (\ref{eq:g0}).}
\end{table}

}
%%%%%%%%%%%%%%%%%%%%%%%%%%%%%%%%%%%%%%%%%%%%%%%%%%%%%%%%%%%%%%%%%%%%%%%%%%%%%%%%%%%%%%%%%%%%%%%%%%%

%%%%%%%%%%%%%%%%%%%%%%%%%%%%%%%%%%%%%%%%%%%%%%%%%%%%%%%%%%%%%%%%%%%%%%%%%%%%%%%%%%%%%%%%%%%%%%%%%%%
%%%%%%%%%%%%%%%%%%%%%%%%%%%%%%%%%%%%%%%%%%%%%%%%%%%%%%%%%%%%%%%%%%%%%%%%%%%%%%%%%%%%%%%%%%%%%%%%%%%
\section{Conclusion}

We examined pair distribution functions and local distributions for the 1D dilute Ising model. 
The thermodynamic properties of the Ising chain with impurities qualitatively differ from the case of a pure Ising chain and depend on the ratio of the exchange constant ($J$) and the impurities interaction parameter ($V$). 
An analysis shows that if $V<|J|$, the phase separation occurs when the system is divided into macroscopic domains consisting only of impurities or only of spins. 
In this case, both the spin and impurity correlation lengths diverge at zero temperature. 
If $V\geq|J|$, the system is frustrated and its entropy does not equal zero at zero temperature. 
While the spin correlation length is always finite, the impurity correlation length tends to infinity 
at half-filling and $V>|J|$. 
If $V\neq|J|$, the specific heat of the dilute Ising chain has a distinct maximum at some critical temperature 
due to the ordering processes in the impurities subsystem. 
We found that this critical temperature is determined by the maximum rate of change of the impurity–spin pairs concentration and is proportional to the modulus of a difference between the exchange constant and the impurities interaction parameter. 
Unlike the dilute Ising chain,  
in the two-dimensional dilute Ising system with mobile impurities and in related models 
\cite{Arora1973,Wortis1974b,Yaldram1993,Khalil1997,Loois2008,Panov2019}, 
the previously considered orderings are the phase transitions at finite temperatures. 
If $V<|J|$ and the impurities concentration $n<n_c$, the spin ordering with the temperature lowering precedes to the phase separation when the domains containing only impurities or only spin centers occur. 
If $n>n_c$, the phase separation appears simultaneously with the spin ordering in the spin domains. 
If $V>|J|$, the diluted spin ordering at $n<n'_c$ is changed at $n>n'_c$ by the charge ordering of impurities 
with the short-range spin order.

To consider the details of the ordering processes, we explored the local distributions. 
The Markov property of the dilute Ising chain allowed us to use the conditional probabilities for the nearest neighbors to obtain the expressions of the pair distribution functions, the probability of any finite sequence and, in particular, of an isolated sequence of repeating blocks. 
The lengths of isolated sequences of repeating blocks $\boldsymbol{\sigma}$ obey the geometric distribution, 
and the probability of the cycle $\boldsymbol{\sigma}$ is the only parameter determining its properties. 
If the mean length of some repeating sequence tends to infinity, 
the dispersion of lengths also tends to infinity, asymptotically as a square of the mean length. 
We showed that the critical behavior of the spin correlation length at $V<|J|$ 
is defined by ferromagnetic sequences for $J>0$ and by antiferromagnetic sequences for $J<0$, 
while the critical behavior of the impurity correlation length is defined by the sequences of impurities at $V<|J|$ 
and by the charge-ordered sequences at half-filling and $V>|J|$.
We concluded that for the dilute Ising chain there are no other repeating sequences 
for which the mean length tends to infinity at zero temperature. 
We found that the non-ordered dilute Ising chain corresponds to the regular Markov chain, while the ordering generates the irregular Markov chain.

%%%%%%%%%%%%%%%%%%%%%%%%%%%%%%%%%%%%%%%%%%%%%%%%%%%%%%%%%%%%%%%%%%%%%%%%%%%%%%%%%%%%%%%%%%%%%%%%%%%
%%%%%%%%%%%%%%%%%%%%%%%%%%%%%%%%%%%%%%%%%%%%%%%%%%%%%%%%%%%%%%%%%%%%%%%%%%%%%%%%%%%%%%%%%%%%%%%%%%%
\section*{Acknowledgments}
This work was supported by Program 211 of the Government of the Russian Federation, Agreement 02.A03.21.0006,
and the Ministry of Education and Science of the Russian Federation, project FEUZ-2020-0054.

%%%%%%%%%%%%%%%%%%%%%%%%%%%%%%%%%%%%%%%%%%%%%%%%%%%%%%%%%%%%%%%%%%%%%%%%%%%%%%%%%%%%%%%%%%%%%%%%%%%
%%%%%%%%%%%%%%%%%%%%%%%%%%%%%%%%%%%%%%%%%%%%%%%%%%%%%%%%%%%%%%%%%%%%%%%%%%%%%%%%%%%%%%%%%%%%%%%%%%%

\end{document}